\documentclass{article}
\usepackage{epsf}
\usepackage{emulateapj}
\def\etal{et~al}

\begin{document}

%\shortauthors{Robertson \& Leiter}
%\shorttitle{BHC Magnetic Moments}
%\slugcomment{Submitted to ApJ Feb. 22, 2001}

\title{Evidence For Intrinsic Magnetic Moments in Black Hole Candidates}

\author{Stanley L. Robertson \altaffilmark{1} and Darryl J. Leiter
\altaffilmark{2}}

\altaffiltext{1}{Physics Dept., Southwestern Oklahoma State
University, Weatherford, OK 73096, USA, roberts@swosu.edu}
\altaffiltext{2}{FSTC, Charlottesville, VA 22901, USA, dleiter@aol.com}

\begin{abstract}

We present evidence that the power law part of the quiescent x-ray emissions of
neutron stars in low mass x-ray binaries is magnetospheric in origin. It
can be very accurately calculated from known rates of
spin and magnetic moments obtained from the $\sim 10^{3 - 4}$
times brighter luminosity at the hard spectral state transition.
This strongly suggests that the spectral state
transition to the low hard state for neutron stars is
a magnetospheric propeller effect.
We test the hypothesis that the similar spectral state switches and
quiescent power law emissions of the black hole candidates might be
magnetospheric effects. In the process we derive proposed magnetic moments and
rates of spin for them and accurately predict their quiescent luminosities.
This constitutes an observational test for the physical realization of event
horizons and suggests that they may not be formed during the gravitational
collapse of ordinary matter.
\end{abstract}


\section{Introduction}
It has recently been claimed that the very low, but nonzero, quiescent
luminosity levels of galactic black hole candidates (GBHC) in low mass x-ray
binary systems (LMXB) constitutes evidence of event horizons (Garcia \etal.
2001). This interpretation of the observations relies on an advection dominated
accretion flow (ADAF) model (Narayan, Garcia \& McClintock 1997)
in which the energy released by accretion is stored
as heat in a radiatively inefficient flow which eventually disappears
through an event horizon. Garcia \etal. (2001) state their belief that
any straightforward explanation of the differences of quiescent luminosities
of GBHC and neutron star (NS) systems, whether based on ADAFs or not, will
require postulating an event horizon in GBHC. In this work we will show that
their quiescent luminosities may have a rather mundane
explanation as magnetospheric emissions.

This surprising assessment arises from the strong broad band
similarities of NS and GBHC, particularly in quiescence
(e.g., Tanaka \& Shibazaki 1996, van der Klis 1994). These similarities
include power law emissions of both NS and GBHC. Photon indexes of $\sim 2$ are
found in both types of object, extending into quiescence. As shown below,
spins of a few hundred Hz, combined with magnetic moments of a few $10^{27}$
G cm$^3$ lead to magnetospheric power law emissions of $10^{32-33}$ erg/s for
NS. This spin-down luminosity is as inevitable in the magnetic NS of LMXB as
it is in ordinary pulsars. But given the similarities
with GBHC, this significantly sharpens the question of how the latter
produce their quiescent emissions. It will be made clear that spins of
a few tens of Hz and magnetic moments of order $10^{29}$ G cm$^3$ would
produce their quiescent luminosities of $10^{30-32}$ erg/s.
Although the existence of objects compact enough to qualify as black hole
candidates is beyond question and the existence of black holes is
accepted by most astrophysicists, it is still observationally unclear
whether event horizons can be physically realized in the collapse of
stellar mass objects. Hence it is still necessary that we be able to
exclude the possibility that GBHC might be intrinsically magnetized
objects before we can say that they truly are black holes. It has recently
been suggested (Mitra 2000) that, within the framework of General Relativity,
trapped surfaces cannot be formed by collapse of physical matter and radiation.
Such a view was also held by Einstein (1939).

In this paper, we observationally test whether event horizons are necessary
for the explanation of the low quiescent x-ray luminosities of GBHC.
Using only the most rudimentary generic features
of strong-field gravity, we consider a model of GBHC with intrinsic magnetic
moments. We assume the magnetic fields to be intrinsic because we require
them to be anchored and co-rotating with the
central object. We find that they also need to be stronger than those
that can be generated via disk dynamos (Livio, Ogilvie \& Pringle 1999) and
more persistent than can be expected with intense accretion onto charged
Kerr-Newman black holes (Punsly 1998). Field strengths of magnitude similar
to those we consider have been implied by the kinetic power of the jets
and synchrotron emissions of the GBHC, GRS 1915+105 (Gliozzi, Bodo \&
Ghisellini 1999, Vadawale, Rao \& Chakrabarti 2001). We assume that the magnetic
field mediates the spectral state switches of NS and also drives the power law
part of the quiescent emissions. We find that the magnetospheric model
accords with many other observed features.

Even the weakly magnetic ($\sim 10^8$ G)
atoll class NS produce significant magnetic stresses on accreting plasma.
At the co-rotation radius (where the Keplerian
orbit frequency matches the magnetosphere spin rate) the magnetic
pressure is $\sim 10^7$ bar. When the inner radius of the accretion
disk expands beyond the co-rotation radius, the
magnetosphere acts as a propeller, cutting off the flow to the
NS surface. This causes a spectral state transition as surface impacts and
soft surface thermal emissions abate (Campana \etal. 1998, Zhang,
Yu \& Zhang, 1998, Cui 1997, Stella, White \& Rosner 1986).
The magnetic propeller effect has been
invoked (Menou \etal. 1999) to explain how an ADAF might be prevented
from reaching the surface of a quiescent NS and producing high NS luminosity.
It is widely believed that ADAFs for NS are stopped by the propeller
effect, otherwise a typical, optically implied, ADAF mass flow rate of
$5\times 10^{15}$ g/s would produce a very non-quiescent luminosity of
$\sim 7\times 10^{35}$ erg/s upon reaching the surface of a NS.
But such a flow rate would at least reach the light cylinder radius of
$\sim 100$ km for an atoll spinning at 500 Hz. If reversed
there, it should then radiate $\sim 5\times 10^{34}$ erg/s,
which is between one and two orders of magnitude larger than the observed
quiescent luminosities of NS. Bisnovatyi-Kogan and Lovelace (2000)
also find that electron heating of an ADAF plasma may lead to
luminosities about 25\% as large as those of a conventional thin disk with
the same accretion rate, which is far beyond quiescent NS luminosity.
These quantitative difficulties with the ADAF
model, the similar spectral state switches of GBHC, and the need to
test the claimed detection of event horizons have provided
the motivation for considering a magnetospheric alternative.

\section{Analysis}
Quiescent x-ray emissions of NS show soft surface thermal and power law
components of roughly the same luminosity
(Campana \etal.2000, Rutledge \etal. 2001).
The magnetospherically driven power law emissions that we consider here
would be the minimum quiescent luminosities for NS.
As we are not accounting for surface thermal emissions, factor of two
disagreements between measured and calculated quiescent luminosities would
be acceptable; indeed even larger variances could occur as a result of
the stochastic nature of the magnetospheric emissions.
As this is being written, no soft thermal emissions have been reported
for quiescent GBHC. Garcia \etal. (2001) report that their quiescent emissions
are consistent with single power laws with photon index $\sim 2$, similar
to those of NS. We shall assume that the power
law emissions are magnetospheric for both NS and GBHC. These
emissions can be calculated as magnetospheric spin-down luminosities using
Becker \& Tr\"{u}mper's (1997) correlation of x-ray luminosity with rate of
loss of rotational energy, $ \dot{E} = 4 \pi^2 I \nu \dot{\nu}$, where $I$
is the moment of inertia of the star and $\nu$ its rate of spin.
We assume that
the energy loss derives from the magnetic torque on an axially aligned rotating
magnetic dipole, which gives $\dot{E} = 32\pi^4 \mu^2 \nu^4/3c^3$,
(Bhattacharya \& Srinivasan 1995)
where $\mu$ is the magnetic moment and $c$ the
speed of light. (Equating these energy loss rates and taking $I = 10^45$ g cm$^2$
provides the standard equation for determining magnetic moments from pulsar
spin-down data $B = 3.2\times 10^{19} (P\dot{P})^{1/2}$ G, with $P=\nu^{-1}$.)
Becker \& Tr\"{u}mper found the x-ray
luminosity to be about $10^{-3}\times 4 \pi^2 10^{45}\nu \dot{\nu}$ erg/s.
Thus the quiescent x-ray luminosity would then be given by :
\begin{equation}
L_q = 10^{-3}\times 10^{45} \frac{32 \pi^4 \mu^2 \nu^4}{3c^3 I} = (4.8\times 10^{30} erg/s)
\frac{\mu_{27}^2 \nu_2^4}{m R_6^2}
\end{equation}Here we have taken the magnetic moment to be $\mu_{27} \times
10^{27}$ G cm$^3$, the spin frequency to be $100 \times \nu_2$ Hz, the
mass of the star as $m$ in M$_\odot$, and the radius as $10^6\times R_6$ cm.

We assume that accretion flow takes place in a geometrically thin,
optically thick accretion disk. Except in
quiescence, the inner disk radius is assumed to be the same as the
magnetospheric radius, $r_m$. This radius is the location at which
the magnetosphere is capable of the appropriate rate of removal of
angular momentum from the inner disk. Mathematically, the result is similar
to that obtained by equating the impact pressure of accreting plasma
to the magnetic pressure. For spherically
symmetric accretion (e.g. Campana \etal. 1998, Equation 1):
\begin{equation}
r_m~=~K{(\frac{\mu_{27}^4}{m \dot{m}_{15}^2})}^{1/7}
\end{equation}where $\dot{m}_{15}$ is the mass accretion rate in
units of $10^{15}$ g/s and K is a scale factor. For spherical accretion,
$K = 233$ km, however, as noted by Campana \etal. (1998), this needs to
be corrected by a factor of $\sim 3$ to account for the difference
between disk accretion and spherically symmetric accretion.
For this reason, we will take $K = 80$ km. This brings the
magnetospheric radii into close agreement with those of an elaborate
model of a gas pressure dominated disk (Ghosh \& Lamb 1992).
(For the present purposes of calculating
luminosities, the extent to which the disk is threaded by the magnetic field of
the central object is unimportant because it does not directly have a large
effect on the accretion energy released for a given inner disk radius.)
$K = 80$ km, is a poorly known parameter and the calculations to
follow are fairly sensitive to its value. Errors in $K$ are reflected in errors
of determination of $\mu_{27}$ and then in calculations of $L_q$.

The spectral state switch takes place when the disk inner radius matches
the co-rotation radius, $r_c$. This can be expressed in terms of the
Keplerian orbit frequency, which is also the spin rate of the star, as:
\begin{equation}
r_c = (70~km){(\frac{m}{\nu_2^2})^{1/3}}
\end{equation}
The light cylinder radius is given by $r_{lc} = (480~ km)/\nu_2$.
When the disk inner radius lies inside the light cylinder, we assume that
the disk luminosity is given by
\begin{equation}
L = \frac {GM\dot{m}}{2r_m}
\end{equation}
Here G is the Newtonian gravitational force law
constant, and $\dot{m}$ the accretion rate.
This corresponds to conversion of half the accretion potential energy to
luminosity via viscous dissipation. This cannot be entirely accurate as
some energy extracted from the rotation of the central object drives the
propeller outflow and may contribute to the
luminosity. In addition, relativistic corrections are needed near
the innermost marginally stable orbit, however, these are unimportant
for radii as large as most co-rotation radii for the objects considered here.
When all of the accreting matter
actually strikes the star, we take the luminosity to be
\begin{equation}
L = \eta \dot{m}c^2
\end{equation}
where $\eta$ is the efficiency of conversion of mass-energy to luminosity,
which should be determined from models of gravitating objects in strong-field
limits. For NS, $\eta = GM/(c^2R) \sim 0.14$ is both customary and
adequate. For GBHC, $\eta$ is a
parameter to be determined from observational data.
Three additional characteristic luminosity levels (Campana \etal. 1998)
are used here. $L_{min}$ corresponds
to the accretion disk having penetrated
inside the co-rotation radius; enough to allow essentially all of
the accreting matter to reach the central object. At this point spectral
softening ends. $L_c$ is the luminosity when
the disk inner radius is only a little beyond the co-rotation radius,
the propeller reverses the inflow and a hard spectrum is produced.
$L_{q,max}$ is the luminosity with disk radius at
the light cylinder. Using $r_c = r_m$ to eliminate $\dot{m}$ and
setting $r = r_c$ in Equations 4 and 5 there follows:
\begin{equation}
L_{min} = (1.4\times 10^{36} erg/s)\eta \mu_{27}^2 \nu_2^{7/3} m^{-5/3}
\end{equation}
and with $r = r_c = r_m$
\begin{equation}
L_c = (1.5 \times 10^{34} erg/s) \mu_{27}^2 {\nu_2}^3 m^{-1}
\end{equation}
Similarly, using $r_m = r_{lc} = r$ yields
\begin{equation}
L_{q,max} = (2.6\times 10^{30} erg/s) \mu_{27}^2 \nu_2^{9/2} m^{1/2}
\end{equation} For NS we shall assume that $m = 1.4$, $R_6 = 1.5$
and $\eta = 0.14$ unless otherwise noted. For GBHC we have used
literature values for mass.

As the magnetic moment, $\mu_{27}$, enters each of Equations 1, 6, 7, 8, it can
be eliminated from ratios of these luminosities, leaving relations involving
only masses and spins. With mass values given, the ratios then yield the spins.
Here we will use Equation 6 or 7 to find $\mu_{27}$, after
which Equation 1 yields $L_q$. Alternatively, if the spin is known from
burst oscillations, pulses or spectral fit determinations of $r_c$,
one only needs one measured luminosity
to enable calculation of the remaining $\mu_{27}$ and $L_q$.

\begin{table*}
\begin{center}
\caption{Calculated and Observed Quiescent Luminosities}
\begin{tabular}{lrrrrrrrr} \hline
Object & m & $L_{min}$ & $L_c$ & $\mu_{27}$ & $\nu_{obs}$ & $\nu_{calc}$ & log (L$_q$) & log (L$_q$) \\
    & M$_\odot$ & $10^{36}$erg/s & $10^{36}$erg/s & Gauss cm$^3$    & Hz      &  Hz~ & erg/s & erg/s \\
    & & & & & obs. & calc. & obs. & calc. \\ \hline
Aql X-1 & 1.4 & 1.2 & 0.4 & 0.47 & 549 & 658 & 32.6  & 32.5  \\
4U 1608-52& 1.4 & 10 &2.9 & 1.0 & 619 & 534 & 33.3 & 33.4 \\
Sax J1808.4-3658& 1.4 & $^a$0.8 & 0.2 & 0.53 & 401 & 426 & 31.8-32.2 & 32 \\
Cen X-4 & 1.4 & 4.4 & 1.1 & 1.1 & & 430 & 32.4 & 32.8 \\
4U 1916-053 & 1.4 & $\sim$14 & 3.2 & 3.7 & 270 & 370 & & 33.0 \\
KS 1731-26 & 1.4 & & 1.8 & 1.0 & 524 & &  & 33.1  \\ %Sunyaev, R. 1990 Soviet Astron. Lett. 16,59
4U 1730-335 & 1.4 & 10 & & 2.5 & $^b$307 & & & 32.9 \\
Cir X-1 & 1.4 & 300 & 14 & 170 & & 35 &  & 32.8 \\ \hline
Cir X-1 & 7 & 300 & 14 & 420 & & 33 &  & 32.6 \\ \hline
GBHC \\
GRS 1124-68 & 5 & 240 & 6.6 & 720 & & 16 & $<32.4$ & 31.9 \\
GS 2023+338 &7 & 1000 & 48 & 470 & & 46 & 33.7 & 33.2 \\
XTE J1550-564 & 7 & $^c$90 & 4.1 & 150 & & 45 & $^d$32.8 & 32.2 \\
GS 2000+25 & 7 & & 0.15 & 160 & & 14 & 30.4 & 30.2\\
GRO J1655-40 & 7 & 31 & 1.0 & 250 & & 19 & 31.3 & 31.2 \\
A0620-00 & 4.9 & 4.5 & 0.14 & 50 &  & 26 & 30.5 & 30.4 \\
Cygnus X-1 & 10 & & 30 & 1260 & & 23 &  & 32.7 \\
GRS 1915+105 & 7 & & 12 & 130 & $^e$67 & & & 32.8 \\ \hline

\end{tabular}
\end{center}
$^a$Corrected to 2.5 kpc (in't Zand \etal. 2000, Dotani \etal. 2000),
$^b$(Fox, D. \etal. 2000), $^c$d = 4 kpc, $^d$(Tomsick \& Kaaret 2001)
$^e$GRS 1915+105 Q $\approx 20$ QPO was stable
for six months and a factor of five luminosity change. \hfill \\
Observed quiescent luminosities are from Garcia \etal. (2001). Observed
spins as reported by Strohmayer (2000). Other table entries are
described in the text and appendix. \\
\end{table*}

\section{Calculations}
Observational data and calculations leading to luminosities shown in
Table 1 are detailed in the appendix.

{\sl Neutron Stars---}\\
Sax J1808.4-3658 is the only NS LMXB to show x-ray pulses.
Observed spins of five other NS were
determined from burst oscillations (Strohmayer 2000). For Aql X-1,
4U 1608-52, and Sax J1808.4-3658, the values of spin and $L_c$
or $L_{min}$ were used
to calculate the values of $\mu_{27}$ shown in Table 1. Spins and
magnetic moments were then
used to obtain their calculated quiescent luminosities. The agreement is
very good, however the identification of $L_{min}$ for Sax J1808.4-3658
is somewhat uncertain (see appendix).
On the other hand, Sax J1808.4-3658 {\it is}
a millisecond pulsar for which we can be reasonably certain that Equation 1
would yield its quiescent power-law luminosity. In this case, we can
invert the process and use $\mu$ determined from Equation 1 to
verify our choices of $L_{min}$ and $L_c$.

An important step in this analysis consists of testing the
validity of the use of luminosity ratios to determine both spins and
magnetic moments. For the first two NS of Table 1, we have
used the ratio $L_{min}/L_c$ to determine spins shown as calculated
values in the table. Both agree with observed values within 20\%.
We used these spins and observed $L_c$ to re-evaluate $\mu_{27}$
and then recalculated values of log($L_q$) of 32.5 and 33.5
for them. If small errors of spin arise from the use of ratios and the values
of $L_{min}$ or $L_c$ are not seriously in error, then
compensating shifts of calculated magnetic moments seem to
leave the  calculated quiescent luminosities relatively
unchanged. 

As no burst oscillations have yet been reported for Cen X-4, we used
the ratio $L_{min}/L_q = 4$ to obtain a spin of 430 Hz and magnetic
moment $\mu_{27} = 1.1$. The calculated quiescent luminosity is
$6\times 10^{32}$ erg/s (log($L_q$) = 32.8), compared to
$2.5\times 10^{32}$ erg/s (log($L_q$) = 32.4) observed. The
agreement between luminosities calculated via the ratio
method and those observed for three NS is good, especially
when one considers that the calculated values were
based on information obtained from luminosities that were more
than $10^3$ larger. This strongly suggests that the
spectral state switch for NS is a magnetospheric effect and
that magnetospheric spin-down accounts for much of the
quiescent luminosity.

Quiescent luminosities for 4U 1916-053, KS 1731-26 and 4U 1730-335
were determined using observed spins and values of $L_c$ or $L_{min}$ estimated
at the endpoints of the spectral state transitions. Their calculated
quiescent luminosities are included here as predicted values.
We note that the magnetic moments that we find here are comparable
to those found for millisecond pulsars with similar rates of spin.

Two entries have been included for Cir X-1. Observed luminosities have
been corrected for a distance of 5.5 kpc rather than
the 10 kpc originally reported (Dower, Bradt \& Morgan 1982).
This also brings its magnetic field ($B = \mu R^{-3}$) into agreement with a
value of $\sim 4\times 10^{10}$ G based
(Iaria \etal. 2001) on an accretion disk
inner radius that was obtained from fitting spectral data. Cir X-1 is,
in many respects, the NS that behaves most like a GBHC. It was
thought to be a GBHC for a long time, until it displayed Type 1 bursts.
It has been clearly identified as a Z source (Shirey, Bradt \& Levine 1999).
It seems to have a spin and magnetic moment more like those of the
GBHCs shown further down in Table 1. The lower spin rate is commensurate with
its `khz' QPOs, which have an onset around 20 Hz. The GBHC properties that
Cir X-1 seems to possess may simply be due to a stronger magnetic field,
but it seems plausible that it may also have the larger mass of a GBHC.
If it has the larger mass, then as a burster it clearly has no event
horizon. If not, then it demonstrates that the magnetic moments may
be involved in the production of the high frequency QPO.
It is of interest, that only its calculated magnetic moment
changes significantly if we suppose it to have different mass. This is also
the case for the GBHC.

{\sl Black Hole Candidates---}\\
For the GBHC, we have three unknown parameters, $\mu_{27}$, $\eta$, and $R_6$ to
determine. We use estimates of the inner disk radius at co-rotation
and Equation 3 to determine the spin frequencies. We use Equation 7
to obtain $\mu_{27}$ and then Equation 6 to evaluate $\eta$.
For GRS 1124-68, GS 2023+338, XTE J1550-564
and GS2000+25, inner disk radii obtained
from spectral fits (see appendix) are 400 km, 224 km, 228 km and 500 km,
respectively. The corresponding values of
$\nu_2$ and  $\mu_{27}$ are shown in Table 1. For the first three objects,
for which reliable values of $L_{min}$ have been found, the
values of $\eta$ were, 0.34, 0.48 and 0.49, respectively. For the
remainder of this work we shall use $\eta = 0.4$ as our best guess.
This choice was used for the second tabular entry for Cir X-1.
We note that this choice is consistent with interpreting the
`ultrasoft' radiations of the GBHC as redshifted surface radiations.
If z is the redshift, it is given by $z = \eta/(1-\eta) = 0.7$, for
$\eta = 0.4$. The surface thermal emissions would then be shifted
to  1.7X lower than rest frame energies and they would appear unusually soft.

At this point, we must also estimate the radii of GBHC before we can calculate
the quiescent luminosity. This choice must be consistent with our
choice of $\eta$. In metric gravity theories, $\eta = 1 - \sqrt{g_{tt}}$,
where $g_{tt}$ is the time component of the spacetime metric. In
standard Schwarzschild coordinates, $g_{tt} = (1 - 2GM/c^2R)$. For
$\eta = 0.4$, the radius of a 7 M$_\odot$ object would be 32 km.
In isotropic coordinates, however, R = 21 km would result. Since the
applicability of either number is in question, we will take R = 20 km
for use here. R enters here as part of the expression of rotational
inertia for a GBHC. Quiescent luminosities are only moderately sensitive to
the value used, but any choice for GBHC is clearly a guess.

Using $R_6 = 2.0$ in Equation 1, we calculate the quiescent luminosities
of GRS 1124-68, GS 2023+338, XTE J1550-564 and GS 2000+25 shown in Table 1.
We use the ratio method with $\eta = 0.4$ and $R_6 = 2$ to obtain quiescent
luminosities for GRO J1655-40 and A0620-00. For the six for which
comparisons can be made, the agreement with
observed values is excellent. The remaining entries in the
table, which lack observed values
for comparison are to be understood as predictions. If we use the ratio
method for XTE J1550-564 rather than the spectrum fit value for $r_c$, we
would find log($L_q$) = 32.1 for $\eta = 0.4$.
An estimated co-rotation radius of 400 km was used to determine the
spin for Cygnus X-1 (see appendix). We identify a QPO frequency of
GRS 1915+105 as its spin frequency. We predict values of $L_{min}$ to
be $1.6\times 10^{38}$ erg/s for GRS 1915+105 and $6.4\times 10^{38}$ erg/s
for Cyg X-1.

\section{Discussion}
The luminosity at the arrest of decline in the light curves of Aql X-1
(Campana \etal. 1998, Fig. 1) and Sax J1808.4-3658 (Gilfanov \etal. 1998)
near the end of an outburst is a factor of $\sim 5$ higher than the final
quiescent luminosity. The luminosity at the arrest for Aql X-1,
$1.2\times 10^{33}$ erg/s, agrees within 25\% of $L_{q,max}$ calculated
using Equation 8 and the spins and magnetic moments obtained
from spectral state switches. Similar factors of 5 to 10 are found for GBHC.
This indicates that during the decline to quiescence, the inner
radius of the accretion disk continues to expand beyond the light
cylinder. If the inner disk remains optically thick, as little as
$\sim 10^{30}$ erg/s emanating from the central star would be capable
of heating it to the instability point and/or ablating it
if nearer than $\sim 10^5$ km. For NS, it seems clear
that the quiescent inner disk must be essentially empty to this radius.
Even if the flow in the inner disk is an ADAF, the flow rate would be
so low that it could contribute little to the
luminosity. It would also be most peculiar for the ADAF luminosity
to depend, as we have shown, on the spin and magnetic
moment of the NS. It is difficult to accept that high flow rate
ADAFs exist for GBHC, but not for NS of similar orbital period.
A more circumspect view is that
the similar quiescent behaviors of GBHC are due to similar magnetospheric
causes.

The picture of LMXB that emerges from this work is that of magnetized
central objects surrounded by accretion disks that have inner radii
dictated either by stability considerations in quiescence or magnetic
stresses at much higher luminosities. Only at very high accretion rates
does the disk push in to the marginally stable orbit. Since many
spectral and timing characteristics observed for GBHC and NS fit well
within this broad picture, we will consider a few of them. These considerations
have gained impetus from the recent discoveries of hard spectral tails for NS
extending beyond 200 keV (e.g., Di Salvo \etal. 2001).
This removes one of the last
of the qualitative spectral differences thought to differentiate NS and GBHC.
This is not to say that there are no spectral differences of NS and GBHC.
There are clear differences that can be attributed to greater mass for
GBHC. For example, Eddington limits are higher and
Keplerian frequencies at the marginally stable orbit
would be much smaller for GBHC. Another major difference is that a prograde
angular momentum at the $\sim 500$ Hz spins of NS in LMXB might
cause the marginally stable orbit radius to be less than that of the star.
Thus the boundary region at the inner disk
could differ between most NS and GBHC, but there are no
presently known differences that can be attributed to an event horizon.

Cyg X-1 is a persistent source that is usually found in a low, hard state, but
occasionally enters a high, soft state with only a factor of 2 or 3 change
of bolometric luminosity. With the inner disk radius at $\sim 400$ km in
the hard state, the disk efficiency ($GM/2c^2r$) is 0.018, compared
to the 0.4 efficiency for accretion to the surface. Hard disk and
ultrasoft surface radiations can contribute equally to the luminosity if
only 0.045 of the accreting matter actually crosses the co-rotation radius.
Thus a spectral state change can occur for very little change of
accretion rate and small change of overall luminosity. Here the
propeller effect provides a clear and simple explanation of the
mysterious spectral pivoting shown by many GBHC. Clearly one
cannot take the ratio of luminosities here as an indicator of spin.
Only if there is some broader measure of spectral hardness that can
be measured over the entire interval from $L_c$ to $L_{min}$ can
reliable results be obtained. Since a substantial increase of radiation
pressure on the inner disk results from small change of accretion
rate, it appears that Cyg X-1 usually stays relatively well balanced
in a relatively low state. If the accretion rate is highly variable
with the inner disk near co-rotation oscillations can result.

Pulsars such as V0332+53, atoll NS, and GBHC all exhibit flickering
in low states. This has a natural explanation as Rayleigh-Taylor
instability shot leakage across the magnetopause. The larger
amplitude flickering for GBHC and pulsars with large magnetic moments
is a result of infall from larger radii. Long time lags at low
Fourier frequencies (Nowak \etal. 1999) imply very slow
propagation speeds ($< 0.01$c) for the luminous plasma disturbances.
Such slow speeds are to be expected for plasma in a magnetosphere.

When well below the Eddington limit, GBHC are much like the atoll NS
examined here. One of the reasons for this might be that the magnetic
stresses at the co-rotation radius are of the same order, $\sim 10^7$ bar
for both. Z class NS have only slightly larger co-rotation radii than atolls,
but have magnetic fields about 10X larger. The magnetic stresses
at their co-rotation radii are much larger, $\sim 10^{8 - 9}$ bar.
Accretion rates near the Eddington limit are needed to push the
magnetopause inside the co-rotation radius.  An optically thick
boundary layer eventually blankets the magnetosphere and obscures
the radiation from the star. In this way the luminosity can decline
while the accretion rate actually increases, producing the familiar
`Z' track in the hardness vs. luminosity diagram. The track is
also seen in color-color diagrams. Portions of such tracks
are occasionally seen in GBHC such as GX 339-4 (Miyamoto \etal.
1991) or GS 2023+338) ($\dot{Z}$ycki, Done \& Smith 1999a, 1999b).
Typical normal branch oscillations $\sim 4 - 10$ Hz are observed for them.
In addition to Z sources being infrequent bursters that have shown
no burst oscillations, we have generally omitted
Z sources in this work because the radiation pressure materially
affects the inner disk in the spectral switch region and changes the
manner in which radius scales with accretion rate.

A superluminal jet was produced by GRS 1915+105 contemporaneous
with the oscillations described for it in the appendix. Chou \&
Tajima (1999) devised a promising model for such pulsed ejections in which a
poloidal magnetic field of unspecified origin is toroidally wound by the inner
disk until it becomes buoyantly unstable and is ejected from the inner disk.
In their model, they artificially stopped the flow through
the event horizon using a pressure of unspecified origin. It is
widely believed that magnetic fields generated in
the accretion disks of black holes can produce
strong poloidal fields threading the event horizon. This seems unlikely,
(Livio, Ogilvie \& Pringle 1999, Punsly 1998) however, because the
same field threads the disk where magnetic pressures can at
most be comparable to the gas pressure. GBHC that possessed intrinsic
magnetic moments could provide the necessary conditions for applicability
of the Chou \& Tajima model. In any event, disk generated
magnetic fields cannot provide a propeller effect that cuts off the
flow into the interior for long times. After the flow cuts off, the
field would decay in about the light crossing time for the original
region of field. The disk would then refill on a slower viscous time scale.
It is also not obvious that cutting off
the flow to an event horizon could produce a spectral state switch.
Considering the high accretion rates involved in the jet production of
GRS 1915+105, it seems unlikely that the inner charge ring of a charged, rapidly
spinning Kerr-Newman black hole (Punsly 1998) could be the source of the
strong magnetic field at the base of its jet.

In our analyses, we have generally used luminosities obtained in the
declining phase of outbursts. During rising phases, there is a
transition through an intermediate state, complete with onset of several
QPO, when the co-rotation radius is traversed. There may be some hysteresis in
luminosities of the spectral state transitions of rising and declining
phases. This sometimes appears in temporal shifting of the
position of Z tracks in hardness vs. luminosity diagrams. Substantial
changes in luminosity for the tracks over weeks are relatively common.
Such behavior may be due to screening of the magnetic fields by
accumulations of accreted matter on the surface. If the field weakens,
it would permit matter to go deeper into the gravitational well and
thus change the luminosity at which some phenomena such as QPO are
observed.

The inner disk radius and the outer point
of disengagement of disk and magnetosphere would be natural boundaries
for the generation of QPO. Except near the co-rotation
radius the disk material would be subjected to either braking or
accelerating torques. These would produce a shear across the co-rotation
region, depending on the extent of magnetic field threading of the disk.
It would not be surprising to have oscillations generated by the shear.
Most of the current models (beat frequency, relativistic precession and
magnetospheric coriolis oscillators) of high frequency QPO
rely on variations of the inner disk radius to
generate the variation of at least one of the high frequency QPO seen in LMXB.
Recent simulations (Kato \etal. 2001) have shown that another high frequency
QPO is generated by reconnection of magnetic field lines at the inner disk
radius. The field lines are toroidally wound by the inner disk until they break.
For both NS and GBHC, the high frequency QPO begin near the spin
frequency and increase to values nearly commensurate with
that of the marginally stable orbit. The onset near the spin
frequency is consistent with our interpretation of mass beginning to
reach the central object when the corotation radius is traversed.
This consistency adds confidence to our use of disk co-rotation
radii determined from spectral fitting as
these co-rotation radii have provided some of our spins.

The hypothesis that GBHC possess magnetic fields can be tested. Searches
should be undertaken for coherent pulses in the vicinity of the calculated
spin frequencies of Table 1. Unusual GBHC burst events should be examined
for burst oscillations. It might also be possible to find
cyclotron resonance features in the range 0.01 - 1 keV
in the spectra of the GBHC. Dynamic
mass determinations for Cir X-1 and XTE J1550-564 are
definitely needed. The properties of XTE J1550-564 and Cir X-1, including
their high frequency QPOs, are strikingly similar. It is highly unlikely
that Cir X-1 possesses a strong magnetic field while XTE J1550-564 does not.
If XTE J1550-564 possesses a similar magnetic field, it should
display a `Z' track in a color-color diagram similar to that of Cir X-1.
The onset of `khz' QPO at frequencies typically within $\sim$ 50\% of
the spin frequency would seem to confirm the spins found here for
Cir X-1 and XTE J1550-564.

\section{Conclusions}
We have shown that the luminosities of the spectral state switches of
NS in LMXB can be used to accurately predict luminosities that are
$\sim 10^{3-4}$ fainter at the light cylinder and in quiescence.
Spin rates and magnetic moments that we have found for
NS are comparable to those found for known millisecond pulsars.
The magnetospheric model on which these calculations rely depends
strongly ($\mu^2 \nu^4$) on magnetic moment and spin of the NS and is
therefore incompatible with ADAF models. This significantly
sharpens the question of how GBHC produce their quiescent spectra.
We have shown that their
quiescent luminosities can be accurately calculated for the
magnetospheric model from spectral state switch luminosities that
are $\sim 10^4$ brighter. In the process, we have found
a reasonably consistent set of GBHC magnetic moments and spins
for eight GBHC and Cir X-1.  We acknowledge
that the parameters, K, $\eta$ and R$_6$ are poorly constrained, but the
values used here certainly seem plausible. Both $\eta$ and R$_6$ are
expected to be mass dependent quantities.

The soft spectral state in this model arises from the surface
of the central object. This requires a paradigm shift away from
supposing that all soft spectral components arise in the disk. We
attribute the `ultrasoft' radiations of GBHC to greater surface redshifts.
It is well known that power law emissions and hard spectra are
generated by accreting pulsars but there is no agreement on how
they do it. Magnetohydrodynamic simulations of the propeller
regime have not yet provided the details.

We have included predictions and suggested tests of our model.
We expect that all GBHC and NS properties can eventually
be understood in terms of a unified magnetospheric model.
If it can be conclusively shown that GBHC have magnetic moments, then
some way will need to be found, within the framework of General Relativity
to prevent the physical realization of the event horizon.
It is the horizon that disconnects external magnetic fields
from generator currents in the interior.
The changes of theoretical interpretation
necessary to eliminate physical occurrences of the horizon might very well
leave the landscape of compact objects, frame dragging, marginally stable orbits
and curved spacetime little changed.

\acknowledgements{}
We thank Rudy Wijnands for pointing out significant errors in an earlier
draft. We are indebted to John Tomsick for sharing information
prior to its publication and we thank Jeroen Homan for useful information.

\appendix{Appendix}
\section{Observational Data}
\textit{Aql X-1}:Campana \etal. (1998) reported spectral hardening beginning at
$L_{min} =  1.2\times 10^{36}$ erg/s, $L_c = 4\times 10^{35}$ erg/s and
complete cessation of the rapid decline at about $1.2\times 10^{33}$ erg/s,
which we identify as $L_{q,max}$.

\textit{4U1608-52}: Spectral hardening began in decline at $L_{min} = 10^{37}$
erg/s (Mitsuda et al. 1989). We take $L_c = 2.9\times 10^{36}$ for the
March 1990 hard spectral state (Yoshida \etal. 1993). Garcia \etal. (2001)
have reported a quiescent luminosity of $L_q = 2\times 10^{33}$ erg/s.

\textit{Sax J1808.4-3658} reached a luminosity level of
$\sim 2.5 \times 10^{36}$ erg/s, (d = 2.5 kpc) declined slowly to about
$8\times 10^{35}$ erg/s and then began a rapid decline. (Gilfanov
et al. 1998, Heindl \& Smith 1998). Very similar decline characteristics
were shown for Aql X-1, for which we identified the
luminosity at the start of rapid decline and spectral hardening as $L_{min}$.
Sax J1808.4-3658 never displayed a soft spectral component. It is conceivable
that it never reached $L_c$, but if so then the 401 Hz pulses would have
to have originated in interaction between the magnetic field and disk rather
than on polar caps. We think it more likely that $L_c$ was exceeded as
surface bursts have been observed for it. A recent analysis of its
1996 outburst (in't Zand \etal. 2001) has also confirmed that burst
oscillations occur at the 401 Hz spin frequency. By analogy with the light
curve of Aql X-1, we take $L_{min} = \sim 8\times 10^{35}$ erg/s (d=2.5 kpc),
as the luminosity at the start of rapid decline. With this choice, we
obtain $\mu_{27} = 0.53$ and calculate a quiescent luminosity of $10^{32}$ erg/s,
in reasonable agreement with observation. We also calculate
$L_c = 1.9\times 10^{35}$ erg/s (for d = 2.5 kpc). There may be a change of trend of the
decline at this luminosity similar to that of Aql X-1, however there is
only one luminosity data point to suggest this. But there is also a
change of trend of pulse amplitude fraction at this level. Pulsations
continued to be observed during the entire decline
phase (Cui, Morgan \& Titarchuk 1998) and well below the luminosity that
we calculate as $L_c$. These observations may be consistent with
having a magnetic axis with significant inclination relative to the
the rotation axis. Whatever the case, it should be clear that
Sax J1808.4-3658 {\it is} a pulsar of known spin. Therefore we use
Equation 1 to determine its magnetic moment, we obtain $\mu_{27} =0.44 -  0.62$
(d = 2.5 kpc). Equation 6 then yields $L_{min} = 0.55 - 1.1\times 10^{36}$ erg/s
compared to $\sim 8\times 10^{35}$ erg/s at the
start of its rapid decline. Although our identifications of $L_{min}$ and
$L_c$ appear to be consistent with the observed spin and quiescent luminosity,
we reiterate that these identifications are based only upon the analogy
with Aql X-1. Sax J1808.4-3658 did not show a soft spectrum. If we are
correct that it exceeded $L_{min}$ without producing a soft spectrum, then
we would not expect it to ever become soft if observed at
higher luminosities in subsequent outbursts.

\textit{Cen X-4}: For Cen X-4, the values of
$L_{min}$ and $L_c$ shown in Table 1 were estimated from the
start of rapid decline on May 23, 1979 and the
arrest of decline of the light curve
on May 28, 1979 (Kaluzienski, Holt \& Swank 1980).

\textit{4U 1916-053}: Boirin \etal. (2000) report $L_c = 3.2\times 10^{36}$
erg/s and a soft state luminosity of $L_{min} = 1.4\times 10^{37}$ erg/s.
The 370 Hz spin calculated for it is an upper limit, as it is not
certain that spectral softening was complete.

\textit{4U 1730-335}:Campana \etal. (1998) report $L_c = 10^{37}$ erg/s at
the start of spectral hardening. Fox \etal. (2000) have found a burst
oscillation frequency of 307 Hz.

\textit{KS 1731-26}We take the reported hard state luminosity
(Sunyaev 1990) to be $L_c = 1.8 \times 10^{36}$ erg/s.

\textit{Cir X-1}: Neither spin rate nor mass are available for the enigmatic
Cir X-1. $L_{\min} = 6.3\times 10^{38}$ erg/s and $L_c = 3\times 10^{37}$ erg/s
can be determined from observations just before and two hours after
a spectral hardening transition on Sept 20-21, 1977
(Dower, Bradt \& E Morgan 1982). 

\textit{ GRS 1124-68} Misra \& Melia (1997)
showed that its inner disk radius increased to about $27 R_s$
(here $R_s = 2GM/c^2$, 400 km for 5 M$_\odot$)
after the spectral state transition.
Ebisawa, et al (1994) give a luminosity of $L_c = 6.6\times 10^{36}$ erg/s for
the low state for the period June 13 - July 23, 1991. $L_{min} = 2.4 \times
10^{38}$ erg/s is the average of the period March 10 - April 2, 1991.
The assumption of $r_c = 400$ km yields $\nu_S = 16$ Hz.

\textit{ GS 2023 + 338} never displayed an ultrasoft spectral component,
but on May 30, 1989 it changed luminosity by a factor of 21 in 10 s,
from $10^{39}$ erg/s to $4.8\times 10^{37}$ erg/s (Tanaka \& Lewin
1995). When the luminosity diminished, the reduction in the
1 - 10 keV band was greater than for higher energies, which hints of a
weak spectral state transition. Thus we take these luminosities to be $L_{min}$ and $L_c$,
respectively. From spectral analysis $\dot{Z}$ycki, Done
and Smith (1997b) found $r_{in} = 25 R_G$ (here $R_G = GM/c^2$,
263 km for $7 M_\odot$) and $35 R_G$ (368 km), respectively, for June 20,
1989 and July 19-20, 1989. Luminosities (0.1 - 300 keV) for these dates
based on spectra of Tanaka (1992) are $1.9\times 10^{37}$ erg/s
and $6.3\times 10^{36}$ erg/s,
respectively. The radii scaled as $L^{-2/9}$ yield values of the co-rotation
radius of 214 km and 234 km. We take the mean of 224 km as the
co-rotation radius and find $\nu = 46$ Hz.
Quiescent luminosities of $6.9\times 10^{33}$ erg/s
(Chen, Shrader \& Livio 1997), $8\times 10^{33}$ erg/s (Tanaka \& Shibazaki
1996), and $1.6\times 10^{33}$ erg/s (Garcia \etal. 1997)
have been reported.

{\sl XTE J1550-564---}Spectral data reported (Sobczak \etal. 2000)
for this GBHC showed rapid
decrease of luminosity from $L_{min} = 2\times 10^{38}$ erg/s
(d = 6 kpc) and large changes
of spectral hardness on Oct. 24, 1998 and again on Mar. 12, 1999. A lower
average luminosity level of $L_c = 9.2\times 10^{36}$ erg/s was reached for
the period April 15 - 25, 1999. High frequency QPO's of 285 Hz have
been reported for XTE J1550-564 (Homan \etal. 2000). Assuming these
to originate with the accretion disk radius near the marginally
stable orbit, a mass of $\sim 7 M_\odot$ is implied.
Wilson \& Done (2001) have found an inner disk radius of $22 GM/c^2$ (228 km
for 7 M$_\odot$), from which we obtain a spin of 45 Hz.
The inner disk radius for disk blackbody spectral fits (Sobczak \etal. 2000)
also show a dramatic increase for the April 15 - 25 period in
accord with our identification of $L_c$. Quiescent luminosity
of $7\times 10^{32}$ erg/s (0.5 - 7 keV, d = 4 kpc) has been reported
by Tomsick \& Kaaret (2001), however this was soon after the outburst ended.

\textit{ GRO J1655-40}: $L_{min}$ was reached at $3.1\times 10^{37}$ erg/s
approximately July 29, 1996 (Mendez, Belloni and van der Klis 1998, Fig. 1).
The rapid decline was arrested at about $10^{36}$ erg/s, which can be taken
to be $L_C$. 

\textit{A0620 00}: Spectral hardening began about 100 days after the start of
the 1975 outburst and continued until interrupted by a reflare. At the start of
spectral hardening we find $L_{min} = 4.5\times 10^{36}$ erg/s (Kuulkers 1998,
see Figs 1 \& 2). At the start of the reflare we find
$L_c \approx 1.4\times 10^{35}$ erg/s, though it is not clear that
spectral hardening was complete.

\textit{ Cygnus X-1} exhibits spectral state switches with very little
change of bolometric luminosity at about $L_c = 3\times 10^{37}$ erg/s.
Done \& $\dot{Z}$ycki (1999) find an average inner disk radius of $27 GM/c^2$
(400 km for 10 M$_\odot$) that changes very little as the
spectrum pivots. We find a corresponding spin frequency of 23 Hz.

\textit{GRS 1915 + 105} displays oscillations with peaks above the
Eddington limit followed by hard states that are lower in luminosity
by a factor of 3 - 6. Belloni et al. (1997) have attributed the intervals
between flares to the time required for the inner disk to refill on
a viscous time scale, which is likely correct. Disk blackbody
spectral fits (Belloni \etal. 1997) show that $r_{in}$ oscillates between
about 20 km and 80 km, but occasionally reaches only 55 km, followed immediately
by another burst. 55 km would be the innermost marginally stable
orbit radius for $7 M_\odot$, which is therefore adopted as the
mass. A 67 Hz QPO has been observed (Remillard \etal. 1997,
Remillard \& Morgan 1998) that is sharp ($Q > 20$) and stable
for factors of 5 luminosity change over six months time.
If this is the spin frequency, a co-rotation radius of 174 km is implied.
The low state between flares reached about $4\times 10^{38}$ erg/s for a radius of
about 80 km. Scaling the luminosity ($\propto r^{-9/2}$) for 174 km, we
find $L_c = 1.2\times 10^{37}$ erg/s, $\mu_{27} = 128$ and we predict
$L_{min} = 1.6\times 10^{38}$ erg/s.

\textit{GS 2000 + 25}: $\dot{Z}$ycki, Done and Smith (1997a) have found a
spectral state switch for GS 2000 + 25. When a strong soft component is
present the disk is highly ionized and iron flourescence
strongly relativistically smeared. The disappearance of the soft component
is accompanied by hardening of the power-law and a decrease of ionization.
Little change of inner disk radius occurs at the transition, but the
subsequent decline of luminosity is accompanied by an increase of inner
disk radius. We identify $L_c = 1.5\times 10^{35}$ erg/s for the
hard state of Dec. 16, 1989. If we use an inner radius of
(50 - 100) $R_g$ (500 - 1000 km with $R_g = GM/c^2$) as
the co-rotation radius (Done, C. personal communication)
we obtain  $\nu_S = 14$ Hz for the smaller radius.

\clearpage

\end{document}